\def\ie{\textit{i.e.\ }}

\def\gsim{~\rlap{$>$}{\lower 1.0ex\hbox{$\sim$}}}
\def\lsim{~\rlap{$<$}{\lower 1.0ex\hbox{$\sim$}}}

\documentclass{iaus}
\usepackage{graphicx}

\title[Extrasolar Planet Interactions]{Extrasolar Planet Interactions}
\author[Barnes \& Greenberg]{Rory Barnes$^1$ \& Richard Greenberg$^1$}
\affiliation{$^1$Lunar and Planetary Lab, University of Arizona, 1629 E. University Blvd., Tucson, AZ, USA \\email: {\tt rory@lpl.arizona.edu}}
\pubyear{2008}
\volume{249}
\pagerange{1-10}
\jname{ EXOPLANETS: Detection, Formation and Dynamics}
\editors{S. Ferraz-Mello, \& Y.S. Sun, eds.?}

\begin{document}

\maketitle

\begin{abstract}
The dynamical interactions of planetary systems may be a clue to their
formation histories. Therefore, the distribution of these interactions 
provides important constraints on models of planet formation. We focus
on each system's apsidal motion and proximity to dynamical
instability. Although only $\sim$25 multiple planet systems have been
discovered to date, our analyses in these terms have revealed several
important features of planetary interactions. 1) Many systems interact
such that they are near the boundary between stability and
instability. 2) Planets tend to form such that at least one planet's
eccentricity periodically drops to near zero. 3)
Mean-motion resonant pairs would be unstable if not for the
resonance. 4) Scattering of approximately equal mass planets is
unlikely to produce the observed distribution of apsidal behavior. 5)
Resonant interactions may be identified through calculating a system's proximity to instability, regardless of knowledge of angles such as
mean longitude and longitude of periastron (\eg GJ 317 b and c are
probably in a 4:1 resonance). These properties of planetary systems
have been identified through calculation of two parameters that
describe the interaction. The apsidal interaction can be quantified by
determining how close a planet is to an apsidal separatrix (a boundary
between qualitatively different types of apsidal oscillations,
\eg libration or circulation of the major axes). This value can be
calculated through short numerical integrations. The proximity to
instability can be measured by comparing the observed orbital elements
to an analytic boundary that describes a type of stability known as
Hill stability. We have set up a website dedicated to presenting the
most up-to-date information on dynamical interactions:
http://www.lpl.arizona.edu/$\sim$rory/research/xsp/dynamics.
\end{abstract}

\section{Introduction}
\label{sec:intro}

One of the most striking differences between known exoplanets and the
giant planets of our Solar System involves the observed orbits:
Observed exoplanets tend to have large eccentricities $e$, and small semi-major
axes $a$, whereas the gas giants of the Solar System have small $e$
and large $a$. Recently, however, it has been shown that the dynamical
interactions in many multiple planet systems (including the Solar
System) show certain features in common (Barnes \& Quinn 2004; Barnes \& Greenberg 2006a [BG06a]; Barnes
\& Greenberg 2006c [BG06c]; Barnes \& Greenberg 2007b [BG07b]). These shared traits
suggest that the character of dynamical interactions (over $10^3$ --
$10^4$ years), rather than the present orbits, may be a more
meaningful constraint on the origins of planetary systems (Barnes \&
Greenberg 2007a [BG07a], BG07b). Considerations of shared dynamical properties have
even resulted in the first successful prediction of the mass and orbit
of an extrasolar planet, HD 74156 d (predicted by Barnes \& Raymond [2004] and Raymond \& Barnes [2005]; found by Bean \etal 2008).

Now with over 200 extra-solar planets known, including $\sim 25$
multi-planet systems, we can look at this population as a whole, with
increasing confidence regarding which common characteristics may be
more than statistical flukes. We have identified commonalities among
planetary interactions that may be key constraints on the origins of
planetary systems.  We have identified parameters that quantify an
interaction's proximity to boundaries between qualitatively different
types of motion. These two types of boundaries are the ``apsidal separatrix''
and the dynamical stability boundary. The apsidal separatrix is the
boundary between different types of apsidal
oscillations, \eg libration and circulation. The stability boundary
separates regions in which all planets are bound to the host star from those in which at least one planet is liable to be ejected. These two parameters
are fixed quantities that do not vary over time, but they constrain
how the systems evolve. Systems tend to lie close to
these two boundaries.

In this chapter we review our derivation of
$\epsilon$, which quantifies proximity to an apsidal separatrix, in $\S$
2. Then we describe $\beta$, which parameterizes a system's proximity
to dynamical instability, in $\S$ 3. In $\S$ 4 we present the observed
distributions of these quantities and use them to constrain formation
models. Finally in $\S$ 5 we draw our general conclusions.

\section{The Apsidal Separatrix}
Eccentricities in multiple planet systems oscillate due to
secular interactions (see \eg Laughlin \& Adams 1999; Rivera \& Lissauer
2000; Stepinski \etal 2000; Michtchenko \& Malhotra 2002; BG06a). Thus a currently observed
eccentricity may not be its average value over the $\sim$10,000 year
secular period. Validation of formation models through the eccentricity
distribution is therefore inadequate; the eccentricity oscillations
are a better description of a multiple-planet system's properties.

We consider analytic and numerical models of planet-planet
interactions to describe the oscillations. The analytic approach is
secular theory which considers long-term averages of forces between
planets. This method was originally developed independently by Laplace
and Lagrange, but see BG06a for a presentation that does not involve
matrix manipulation. We compare the secular solutions
to $N$-body numerical models that solve the force equations directly, and are
therefore arbitrarily accurate, for an sufficiently small timestep, using the MERCURY6 code (Chambers \etal 1999).

In general, secular theory predicts that $e$'s and
$\Delta\varpi$'s (the difference between two longitudes of periastron, $\varpi$) oscillate and that $a$ is constant. Therefore,
conservation of angular momentum requires that as one planet's
eccentricity drops, another must rise (as $e$ increases, orbital angular momentum decreases). The type of oscillation depends on initial
conditions. If $\Delta\varpi$ oscillates about 0, the system
is experiencing ``aligned libration.'' If $\Delta\varpi$ librates about
$\pi$, then the system is undergoing ``anti-aligned libration.'' If
$\Delta\varpi$ oscillates through $2\pi$ then the apsides undergo
``circulation''.   The boundaries between these qualitatively different types of behavior are known as ``separatrices''.

\begin{figure}
\centering
\includegraphics[height=8cm]{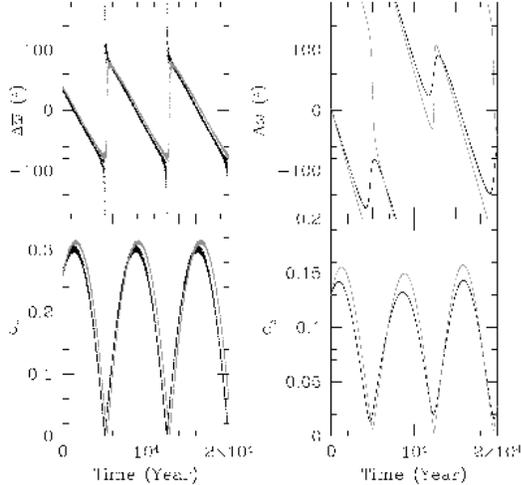}
\caption{Examples of near-separatrix motion in planetary systems. \textit{Left Panels:} A libration-circulation separatrix. Two possible evolutions of $\upsilon$ And c and d (the middle and outer planet of the system) assuming the current orbits. The black points are the system from Butler \etal (2006), the gray from Ford \etal (2005). Although the best-fit orbits in these two cases are very similar, they result in qualitatively different types of evolution of $\Delta\varpi$ (top): The older data predict aligned libration, whereas the updated data predict circulation. Note that the evolution of $e_c$ is similar in both cases, and periodically reaches near-zero values (bottom). \textit{Right Panels:} A circulation-mode separatrix. HD 69830 c and d evolve near the circulation-mode separatrix. The black data are from Lovis \etal (2006), and the gray data are for a fictitious system in which the inner planet's, b's, eccentricity was changed from 0.1 to 0.15. In the first $2 \times 10^4$ years, the actual $\Delta\varpi$ undergoes 1 complete rotation through 360$^o$, but in the fictitious system, $\Delta\varpi$ undergoes 2 complete circulations (top). We again see that the middle planet's eccentricity periodically drops to near-zero values in both cases (bottom).}
\label{fig:separatrix}
\end{figure}

Recently it has been noted that many systems lie near an ``apsidal
separatrix'' (Ford \etal 2005; BG06a, BG06c). For systems of just two planets, the
apsidal separatrix can only separate circulation and libration. This
type of separatrix is a ``libration-circulation separatrix''. An
example of the libration-circulation separatrix is shown in the left
panels of Fig.\ \ref{fig:separatrix}.

In systems of more than two planets, things get more complicated. In
addition to the libration-circulation separatrix, the system may
interact with different numbers of rotations of $\Delta\varpi$ through
$360^o$ during one eccentricity oscillation. The boundary between
interactions with different numbers of circulations in one
eccentricity cycle is a ``circulation-mode separatrix'', and an
example is shown in the right panels of Fig.\ \ref{fig:separatrix}. Note that the gray curve in the top right panel of this figure circulates once and librates once during one eccentricity cycle, whereas the black curve circulates twice.

For an interaction to lie near a separatrix, the amplitude of
eccentricity oscillations is generally two orders of magnitude or
more. Since $0 \le e < 1$ for bound planets, this means that at least
one planet in near-separatrix interactions (both libration-circulation
and circulation-mode) periodically is on a nearly circular orbit. The
proximity to the separatrix can be parameterized as
\begin{equation}
\label{eq:epsilon}
\epsilon \equiv \frac{2[\textrm{min}(\sqrt{x^2 + y^2})]}{(x_{max} -
x_{min}) + (y_{max} - y_{min})},
\end{equation}
where $x
\equiv e_1e_2\sin (\Delta\varpi)$; $y \equiv e_1e_2\cos (\Delta\varpi)$, and the subscripts $min$ and $max$ refer to the extreme values these variables attain over a complete secular cycle (BG06c). (Note that $\epsilon$ is a
fixed quantity even as $e$ and $\Delta\varpi$ change periodically.)
When $\epsilon = 0$ the pair is on an apsidal separatrix, and
one eccentricity periodically reaches zero. Note that in the case of mean motion resonant interactions, $\epsilon$ may not necessarily reveal how close an interaction is to an apsidal separatrix (BG06c).

\section{The Dynamical Stability Limit}
\begin{figure}
\centering
\includegraphics[height=12cm]{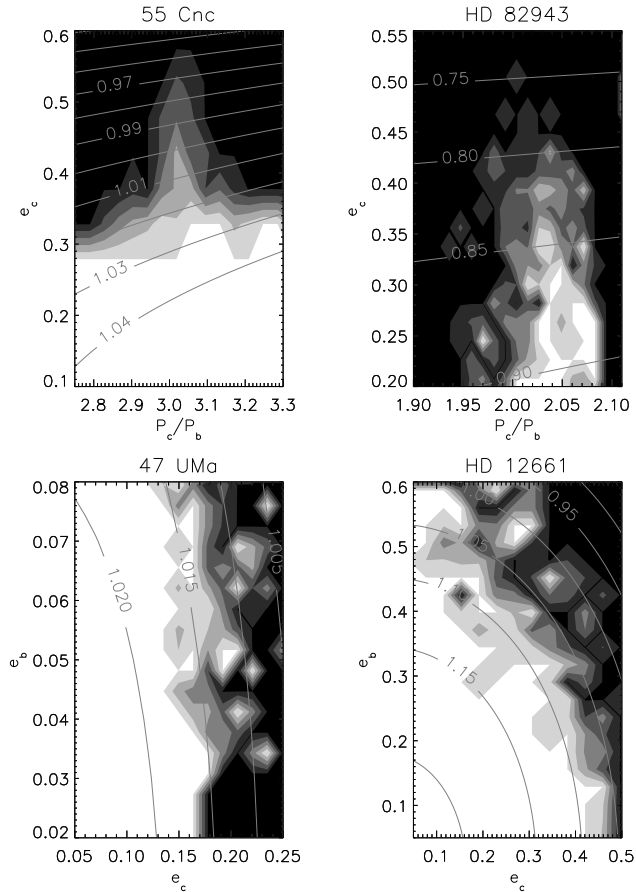}
\caption{Lagrange stability boundary in relation to the Hill stability boundary for four exoplanetary systems. White regions represent bins in which all configurations were stable, black bins contained no stable configurations, darker shades of gray correspond to regions in which the fraction of stable simulations were smaller (\eg Barnes \& Quinn 2004). The curves represent contour lines of $\beta$. Contour lines follow the shape of the Lagrange stability boundary, except in resonance, where the Lagrange stability region is extended to lower values of $\beta$.}
\label{fig:stability}
\end{figure}

Observational uncertainties in
the orbits of exoplanets often include regions of dynamical instability in which one or
more planets would be ejected within 1 Myr, \ie the system is
dynamically ``packed'' (\eg Barnes \& Quinn 2001, 2004; Go\'zdziewski
2002; Kiseleva-Eggleton 2002; \'Erdi \etal 2004; BG06b, BG07b).  This
timescale appears to be long enough to identify nearly all unstable
configurations (Barnes \& Quinn 2004). Representative cases of these
investigations are shown in Fig.\ \ref{fig:stability}. The shading in
this figure correlates with the likelihood of stability: White regions
are stable, black unstable, and grays contain a mix of stable and unstable orbits. In these cases we define stability in the ``Lagrange'' sense for $\sim 10^6$ orbits: Escapes and exchanges of planets are forbidden. 

The preponderance of systems that lie near the stability limit
naturally raises the question of where, exactly, in orbital element space, the
stability boundary lies. Surprisingly, a general analytical expression
for the location of the stability boundary is lacking, even though
it has been the goal of considerable research (\eg Szebehely \& McKenzie
1981; Wisdom 1982; Marchal \& Bozis 1982; Gladman 1993; Laskar 1990;
Holman \& Wiegert 1999; Sosnitskii 1999; David \etal 2003; Cuntz
\etal 2007). These investigations all made significant progress toward defining a stability boundary, but none found it.

Given that many planetary systems lie near the dynamical
stability limit, we have considered how to quantify the proximity to
stability (BG06b, BG07b). We will focus on
one formulation by Marchal \& Bozis (1982; see also Gladman 1993) that
considers ``Hill stability''. Hill stability only requires the
ordering of the planets to remain constant for all time; the outer
planet may escape to infinity.

A coplanar, two-planet system, outside of a resonance is Hill stable if 
\begin{equation}
\label{eq:exact}
-\frac{2M}{G^2M_*^3}c^2h > 1 + 3^{4/3}\frac{m_1m_2}{m_3^{2/3}(m_1+m_2)^{4/3}} -
 \frac{m_1m_2(11m_1 + 7m_2)}{3m_3(m_1+m_2)^2} + ...,
\end{equation}
where $M$ is the total mass of the system, $m_1$ is the mass of the
more massive planet, $m_2$ is the mass of the less massive planet,
$m_3$ is the mass of the star, $G$ is the gravitational constant, $M_*
= m_1m_2 + m_1m_3 + m_2m_3$, $c$ is the total angular momentum of the
system, and $h$ is the energy (Marchal \& Bozis 1982). Here $c$ and $h$ must be calculated in barycentric coordinates. If a given
three-body system satisfies the inequality in Eq.\ (\ref{eq:exact}),
then the system is Hill stable. If this inequality is not satisfied,
then the system may or may not be Hill stable. In this inequality, the
left-hand side is a function of the orbits, but the right-hand side is
only a function of the masses. This approach for identifying stable orbits is fundamentally
different from other common techniques for determining stability which
exploit resonance overlaps (Wisdom 1982; Quillen \& Faber 2006),
chaotic diffusion (Laskar 1990; Pepe \etal 2007), fast Lyapunov
indicators (Froeschl\'e \etal 1997; S\'andor \etal 2007), or periodic
orbits (Voyatzis \& Hadjidemetriou 2006; Hadjidemetriou 2006).

The ratio of the two sides of the inequality in Eq.\ (\ref{eq:exact})
(left-hand side over right-hand side) allows a quantification of a
system's proximity to the Hill stability boundary. We define this
ratio as $\beta$. If $\beta < 1$ a system's Hill stability is unknown,
if $\beta > 1$, Hill stability is guaranteed, if $\beta = 1$, the
system is on the boundary. Note that for any system of two planets outside of resonance with fixed mass, the locus of Hill-stable orbits can be easily calculated.

Fig.\ \ref{fig:stability} includes contour lines of constant $\beta$
for the four systems considered. For the bottom two cases (systems
with no mean motion resonance), the $\beta = 1$ lines are near the
transition from Lagrange stable orbits (white) to unstable
(black). These two cases therefore show that when $\beta \gsim 1$, a system is not
only Hill stable, it is also Lagrange stable.

However, in the top two panels of Fig.\ \ref{fig:stability}, the $\beta = 1$
contour cuts through a swath of stability located at resonance. This
qualitatively different behavior is no surprise because the criterion for Hill stability, Eq.\ (\ref{eq:exact}),
does not apply to systems in resonance. Note however that in the top
left panel, outside of resonance, the $\beta$ contours do follow the
Lagrange stability boundary, again at values near 1. In the top right, we see that all $\beta$ values are less than
1, hence there are no non-resonant, Lagrange stable configurations in
the region of orbital element space considered. 

\section{Dynamical Properties of Exoplanets} 
\begin{figure}
\centering
\includegraphics[height=12cm]{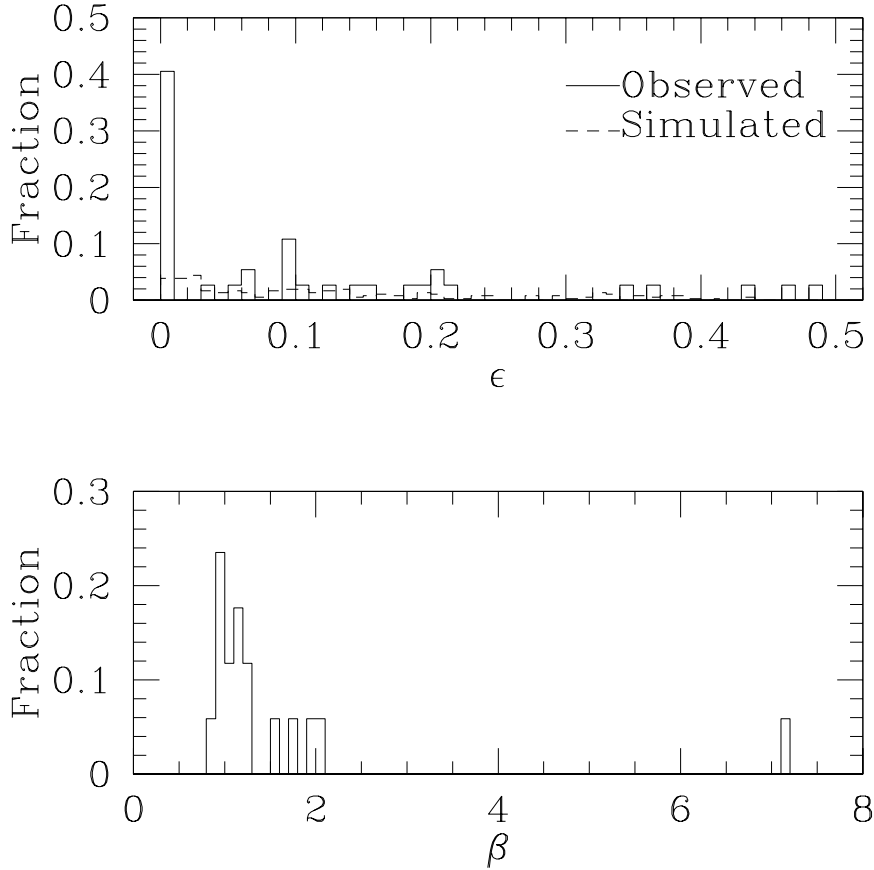}
\caption{\textit{Top Panel:} Distributions of $\epsilon$. The solid line is the observed distribution, the dashed line is that predicted by a model involving the ejection of an additional Jupiter mass planet (Ford \etal 2005; BG07a). \textit{Bottom Panel:} Observed distribution of $\beta$ in known two-planet systems. Note that we may only calculate $\beta$ for two-planet systems.}
\label{fig:epsbeta}
\end{figure}

In this section we will tabulate the observed distributions of $\epsilon$ and $\beta$ and interpret these distributions in the context of models of planet formation. Observed extrasolar planets tend to lie near the apsidal
separatrix. In the top panel of Fig.\ \ref{fig:epsbeta}, we plot the observed
distribution of $\epsilon$ values (solid line). These values were calculated through an
$N$-body integration that required energy conservation to be better
than 1 part in $10^4$ with a symplectic integration technique. Over
40\% of known systems have $\epsilon < 0.01$, including
two of the three giant-planet pairs in the Solar System.

We tabulate the dynamical properties of multi-planet systems in Table
1\footnote{see http://www.lpl.arizona.edu/$\sim$rory/research/xsp/dynamics for an up-to-date list of these properties}, where we also list the sources of the initial conditions we
used. In this table AM stands for ``apsidal motion'' and the
possibilities are circulation (C), aligned libration (A), or
anti-aligned libration (AA). The MMR column lists the resonance, if
applicable. The proximities to the apsidal separatrix, $\epsilon$, are
the values from the literature (BG06c), as are proximities to the Hill
stability boundary, $\beta$ (BG07b). The ``Class'' distinguishes
orbits whose evolution is dominated by tidal (T), resonant (R) or
secular (S) interactions. Table 1 includes the dynamical properties of
the giant planets in our Solar System for comparison.

Given the strong tendency for systems to be near the apsidal separatrix, it is natural to wonder how systems formed that way. The apsidal behavior is tied to the eccentricities. Considerable work by previous researchers has focused on reproducing the distribution of eccentricity values. The
two most studied processes are ``planet-planet scattering'', \ie
gravitational encounters between planets (\eg Rasio \& Ford 1996;
Weidenschilling \& Marzari 1996; Lin \& Ida 1997; Ford \etal 2001;
Marzari \& Weidenschilling 2002), and torquing
by remnant protoplanetary disks that pumps up eccentricities (\eg
Artymowicz 1992; Boss 2000; Papaloizou \etal 2001; Chiang
\& Murray 2002; Goldreich \& Sari 2003; D'Angelo \etal
2006). Simulations of planet-planet scattering can produce
eccentricity distributions that are similar to the observed
distribution (Ford \etal 2001), but models of disk torquing tend to
produce eccentricities that are too low (D'Angelo \etal 2006).

Ford \etal (2005) were the first to consider how scattering may lead to near-separatrix behavior. They showed that the ejection of a hypothetical additional planet could produce the observed near-separatrix behavior among the planets orbiting $\upsilon$ And. In their
model three planets, initially on circular orbits, formed too close together to be stable. Scattering between two planets
results in one of them being ejected. The third planet, orbiting at a
safe distance from the scattering planets, remained on a circular
orbit. The planet that remained bound to the star after the ejection
received an impulsive kick and its eccentricity quickly jumped to a
relatively large value. This sudden change created a new ``initial
condition'' for the secular interaction. The planet that did not
partake in the scattering then began a new secular evolution, but
since apsidal behavior is periodic, that planet's eccentricity would
return to zero and the motion was near-separatrix.

However, Ford \etal only considered one such unstable case, and it was
unclear how likely such a scenario was. The observed $\epsilon$ statistics (solid line in the top panel of Fig.\
\ref{fig:epsbeta}) place a strong constraint on planet-planet scattering
models. BG07a considered $\sim$ 400 hypothetical systems in order to provide a
statistical test of the planet-planet scattering model's likelihood to
reproduce the observed apsidal behavior of extrasolar planetary
systems. In the main part of that experiment, we considered systems
in which all orbital elements were initially the same except for the
mean longitude of the middle planet (of a three-planet system) which was shifted by 1
degree for each successive case (\ie 360 simulations). Stability is most likely independent of
this parameter (see Eq.\ \ref{eq:exact}), therefore there should
be no correlation between initial mean longitude and
$\epsilon$. The resulting distribution of $\epsilon$ is shown in the top panel of Fig.\
\ref{fig:epsbeta} by the dashed line. This scattering (dashed line) produced one-tenth the fraction of near-separatrix
($\epsilon < 0.01$) cases as is observed (solid line).

The planet-planet scattering model had done a reasonable job of reproducing
the observed eccentricity distribution (Ford \etal 2001), but it may be a challenge for it to reproduce the observed apsidal behavior (BG07a). BG07a noted that the path to small $\epsilon$
values seemed quite constrained: The scattered planet needed to be
ejected immediately. This common scattering route to near-separatrix motion led BG07a to suggest that the probability of near-separatrix motion could be enhanced if the perturber is relatively small and/or on an initially eccentric orbit. Such a ``rogue
planet'' may have a better chance of being
removed from a planetary system after one encounter with a giant
planet, but this model will require further testing. The actual distribution of $\epsilon$ must be reproduced by any model of planet formation.  

Next we discuss the implications of the $\beta$ distribution, that is, how close a system is to instability. We tabulate $\beta$ (Table 1) values to produce another
distribution function that describes the range of dynamical properties
of planetary systems. The observed distribution of $\beta$ is plotted in
the bottom panel of Fig.\ \ref{fig:epsbeta}. Most systems have $\beta \sim 1$, nearly all have $\beta < 2$, and the
exception is HD 217107 ($\beta > 7$). This clustering of systems near the Hill
stability limit suggests that many systems form near the edge of
dynamical stability.

In Fig.\ \ref{fig:epsbeta} and Table 1 we see that many systems in
fact have $\beta < 1$. These systems are all in mean motion resonance;
the interaction is stabilized by the resonance. In fact, HD 108874 is
the only known resonance system to have $\beta > 1$. Therefore it
seems that resonances tend to form such that $\beta < 1$.

\medskip
\begin{center}
Table 1 Dynamical Properties of Multiple Planet Systems
\end{center}
\begin{tabular}{cccccccl}
\hline
\noalign{\smallskip}
\label{tab:props}
System & Pair & MMR & AM & $\epsilon$ & $\beta$ & Class & Reference\\
\noalign{\smallskip}
\hline
\noalign{\smallskip}
47 UMa & b-c & - & C$^a$ & 0 & 1.025 & S & Butler \etal (2006)\\
55 Cnc & e-b & - & C & 0.067 & - & T & Butler \etal (2006)\\
 & b-c & 3:1 & C & 0.11 & - & R\\
 & c-d & - & C & 0.158 & - & S\\ 
GJ 317 & b-c & 4:1? & ?$^b$ & - & 0.98 & R? & Johnson \etal (2007)\\
GJ 876 & d-c & - & C$^a$ & 0 & - & T & Butler \etal (2006)\\ 
 & c-b & 2:1 & A & 0.34 & - & R\\
Gl 581 & b-c & - & C & 0.15 & - & T & Udry \etal (2007)\\
 & c-d & - & C & 0.20 & - & T\\ 
HD 12661 & b-c & - & C & 0.003 & 1.199 & S & Butler \etal (2006)\\
HD 37124 & b-c & - & C & 0.009 & - & S & Butler \etal (2006)\\
 & c-d & - & A & 0.096 & - & S\\
HD 38529 & b-c & - & C & 0.44 & 2.070 & S & Butler \etal (2006)\\
HD 69830 & b-c & - & C & 0.095 & - & T & Lovis \etal (2006)\\
 & c-d & - & C & 0.04 & - & S\\
HD 74156$^c$ & b-c & - & C & 0.36 & 1.542 & S & Butler \etal (2006)\\
HD 73526 & b-c & 2:1 & AA & 0.006 & 0.982 & R & Butler \etal (2006)\\
HD 82943 & b-c & 2:1 & C & 0.004 & 0.946 & R & Butler \etal (2006)\\
HD 128311 & b-c & 2:1 & C & 0.091 & 0.968 & R & Butler \etal (2006)\\
HD 108874 & b-c & 4:1 & C/AA$^d$ & 0.2 & 1.107 & R & Butler \etal (2006)\\
HD 155358 & b-c & - & AA & 0.21 & 1.043 & S & Cochran \etal (2007)\\ 
HD 168443 & b-c & - & C & 0.22 & 1.939 & S & Butler \etal (2006)\\
HD 169830 & b-c & - & C & 0.33 & 1.280 & S & Butler \etal (2006)\\
HD 190360 & c-b & - & C & 0.38 & 1.701 & T & Butler \etal (2006)\\
HD 202206 & b-c & 5:1 & C & 0.096 & 0.883 & R & Butler \etal (2006)\\
HD 217107 & b-c & - & C & 0.46 & 7.191 & T & Butler \etal (2006)\\
Hip 14810 & b-c & - & AA & 0.05 & 1.202 & T & Wright \etal (2007)\\
SS & J-S & - & C & 0.19 & - & S & JPL\\
 & S-U & - & C & 0.006 & - & S\\
 & U-N & - & C & 0.004 & - & S\\ 
$\mu$ Ara & c-d & - & C & 0.002 & - & T & Pepe \etal (2007)\\
 & d-b & 2:1 & C & 0.003 & - & R\\
 & b-e & - & C & 0.13 & - & S\\
$\upsilon$ And & b-c & - & C & $1.8\times 10^{-4}$ & - & T & Butler \etal (2006)\\ 
 & c-d & - & C & $2.8\times 10^{-4}$ & - & S\\
\hline
\end{tabular}\\
\noindent
$^a$ The current eccentricity of one planet is 0, placing the pair on an apsidal separatrix.\\
$^b$ The values of mean longitude and longitude of periastron are unknown.\\
$^c$ These values do not incorporate the new planet Bean \etal (2007)\\
$^d$ This pair alternates between circulation and anti-aligned libration.\\

\bigskip

One noteworthy system in this context is GJ 317 (Johnson \etal 2007),
with a $\beta$ value of 0.98 (see Table 1). This value is only
permissible if the system is in a mean motion resonance, and, indeed,
the ratio of the periods is near a whole number (4.02). Usually a mean
motion resonance is identified by calculating the ``resonant
argument'', which depends on mean longitude and longitude of periastron
(see, \eg Murray \& Dermott 1999). However, in the case of GJ 317,
these two angles are unknown, and, hence, the resonant argument is
also unknown. Nonetheless, the system must be in resonance (barring
substantial revisions in the masses, semi-major axes and
eccentricities of the planets) based on our stability
analysis. Therefore consideration of proximity to instability in terms
of the Hill stability boundary may provide an alternative method for
identifying mean motion resonances. As the orbital parameters of the
two planets in GJ 317 are refined, we predict that they will, in fact,
be found to be in resonance.

\section{Conclusions}
\label{sec:conclusions}
We have described here new approaches for parameterizing the dynamical interactions of extrasolar planet systems. 
Many systems interact such that they are near the
stability limit and the apsidal separatrix.  Most mean-motion
resonance interactions would be unstable if not for the resonance, a
feature which may be used to identify resonant interactions (\eg GJ
317 b and c). The distribution of $\epsilon$ shows that the
scattering of approximately equal mass planets is unlikely to produce
the observed distribution of apsidal behavior. The distribution of $\beta$ values shows that many systems formed near the limit of dynamical stability.

These results demonstrate the benefits of the consideration of the
dynamical interactions of multiple planet systems, which may constrain
models of planet formation. For example, as we discussed above, the
proximity of so many systems to an apsidal separatrix ($\epsilon = 0$)
suggests that ``rogue planets'' may have played a major role in
scattering planets to higher eccentricities. The proximity of so many
systems to the stability limit ($\beta = 1$) suggests many systems
form in densely packed configurations. Moreover, consideration of
planetary systems in these terms has revealed that our Solar System
shares dynamical traits with the known multiple planet
systems. Perhaps constraining planet formation models through
dynamical properties, rather than observed orbital elements, will lead
to a universal model of planet formation.

About half of planetary systems are multiple (Wright \etal 2007),
predictions of additional companions are being borne out (Barnes \& Raymond 2004; Raymond \&
Barnes 2005; Bean \etal 2008), and the current distribution of planet
masses suggest there will be many planets with a mass equal to that of
Saturn or less (Marcy \etal 2005), \ie below current detection limits. These three observations imply many
multiple planet systems will be detected in the future. Hence
characterizing extrasolar planet interactions will be a critical
aspect of the study of planet formation for the foreseeable future.

\end{document}